\documentclass[11pt]{article}

\usepackage{charter}

\usepackage{isolatin1}
\usepackage{a4wide}
\usepackage{graphicx}
\usepackage{pst-node}

\title{An Application of Rational Trees in a Logic Programming Interpreter for a Procedural Language}

\author{%
    Manuel Carro \\%
    C.S. Dept., Technical University of Madrid \\%
    \texttt{mcarro@fi.upm.es}%
}

\begin{document}

\maketitle

\begin{abstract}
  \noindent
  We describe here a simple application of rational trees to the
  implementation of an interpreter for a procedural language written
  in a logic programming language.  This is possible in languages
  designed to support rational trees (such as Prolog II and its
  descendants), but also in traditional Prolog, whose data structures
  are initially based on Herbrand terms, but in which implementations
  often omit the \emph{occurs check} needed to avoid the creation of
  infinite data structures.

  \noindent
  We provide code implementing two interpreters, one of which needs
  non-occurs-check unification, which makes it faster (and more
  economic).  We provide experimental data supporting this, and we
  argue that rational trees are interesting enough as to receive
  thorough support inside the language.
\end{abstract}

\section{Introduction}
\label{sec:intro}

Occurs check, i.e., the need to verify whether a variable $X$ occurs
into a term $T$ prior to performing the unification $X = T$ (and
failing if that is the case), is needed in order to provide soundness
to logic programs based on first order logic.  The need to check this
can appears in general, when two terms have to be unified, as
equations of the above form can be generated while unifying them.  For
example, solving
\begin{displaymath}
 f(X, X) = f(g(Y), Y) 
\end{displaymath}
can be reduced to
\begin{eqnarray*}
    X & = & g(Y) \\
    X & = & Y
\end{eqnarray*}
which in turn boils down to $X = g(X)$.  The relevance of occurs check
when reasoning on programs supposedly executing over the Herbrand
domain can be seen from the following examples, taken
from~\cite{crnogorac96comparison}:

\paragraph{Correspondence with First-Order Semantics}

Consider the statement $\exists z : p(z, z)$.  It is not a logical
consequence of $\forall x \exists y : p(x, y)$ (this is easy to see by
taking $p(x, y)$ as, e.g., $y = x + 1$ in the domain of the natural
numbers).  However, in the Prolog program

\begin{verbatim}
q:- p(X, X).
p(X, Y):- Y = s(X).
\end{verbatim}

\noindent
the query \verb+?- q.+ will succeed if occurs-check is not performed
(as it is the default case\footnote{It is to be noted that the ISO
  Prolog standard~\cite{iso-prolog} does require the presence of a
  builtin (\texttt{unify\_with\_occurs\_check/2}) able to perform
  unification with occurs check.} with many current Prolog compilers
and interpreters).  This means that Prolog without occurs-check is not
adequate as a theorem prover for first-order logic.

\paragraph{Program Transformation}

Some simple program transformations rely on occurs-check being
performed, or else the transformed program can behave differently from
the original one.  For example, the clause
\begin{verbatim}
q(X, Y, U, V):- X = Y, X = a, U = f(X), V = f(Y).
\end{verbatim}
can be trivially transformed into
\begin{verbatim}
q'(X, Y, f(X), f(Y)):- X = Y, X = a.
\end{verbatim}
The call
\begin{verbatim}
?- q(X, Y, X, Y).
\end{verbatim}
finishes with failure in all Prolog
systems, but the call
\begin{verbatim}
?- q'(X, Y, X, Y).
\end{verbatim}
does not terminate in some Prolog systems, depending on how
unification of circular terms is handled.\footnote{In particular, we
  have verified that Yap-5.4.1 and SWI-5.2.7 did not terminate, but
  SICStus and Ciao Prolog (which share an original common code base),
  did.  Of course, newer versions of Yap and SWI may have already
  filled in this gap.}

\bigskip

Besides ensuring soundness, the assumption of the existence of occurs
check makes it possible to carry around unification equations in
\emph{solved form}, i.e., in the form of a set of equations of the
form $Var_i = Term_i$ where no $Var_i$ occurs inside any $Term_j$.
This permits, for example, projecting equations over variables (i.e.,
what a toplevel does to print answers) in a relatively straightforward
way.
Occurs-check is however not without a cost: checking for occurrences
is, using a naïve algorithm, quadratic with respect to the size of the
terms to be unified.  More refined algorithms can reduce the
complexity to
$O(n)$~\cite{MaMo76,MartelliMontanari82,PatersonWegman78}, but
requiring large memory space or with a linear constant too high.  This
has somehow prevented the widespread use of occurs-check unification.
Quoting~\cite{ALP93:occurs_check}:

\begin{quote}
\textit{An extreme example is \texttt{append/3} for difference lists.
  Without the occurs check, the cost is $O(1)$; with the occurs check,
  the cost is $O(n)$ on the size of the terms (in other words, it gets
  the same cost as regular \texttt{append/3}).  Normally, the
  programmer knows that the occurs check is not needed, so this is a
  big win.}  \hfill \emph{[P. Ludemann]}
\end{quote}

Benchmarking of Prolog systems which have included support for occurs
check reports modest overheads in the execution (around 10\%-15\% of
global speed loss, with some studies reporting only 5\%
degradation~\cite{dreussen89:risc_prolog}).  While this might be seen
as a relatively small price to pay, one might wonder if is worth
paying it every time a unification takes place, especially taking into
account that the programmer often knows where cyclic terms can be
generated.  Moreover, cyclic terms bring several useful
characteristics, both in expressiveness and in performance, and we
argue that they should be properly supported.  Note that leaving the
decision of performing or not occurs check to some analysis is not
satisfactory, as the programmer may require it or not, so some sort of
annotation guiding the compiler will be needed in that case.  The
analyzer might otherwise decide to include occurs check unification
where they are not intended.  In fact, as we will see, deciding where
these are necessary is not straightforward, and different applications
can be at odds with respect to the required behavior of, e.g., some
library predicates.

Not using occurs check at all changes the term universe, which must
therefore be given a proper semantics with respect to unification and,
if needed, with respect to \emph{difference} constraints.  This has
been studied
elsewhere~\cite{colmerauer-equations,Colmerauer82,jaffar86:infinite_tree_lp},
and the resulting term universe has been named \emph{rational trees},
as opposed to \emph{Herbrand terms}.  Rational trees are part of the
design of, for example, the Marseilles family of Prolog
systems~\cite{Prologii,colmerauer-prologIII,prologiv-manual,colmerauer-specifications,benhamou-prolog4}.
Even if infinite, in the sense that there are paths of infinite
length, rational trees can be finitely represented and used to
represent cyclic data structures.

There are practical reasons to allow the creation of cyclic terms.  In
practice, it seems that using rational trees does not need more
resources than using Herbrand terms, and the necessary algorithms (not
only unification, but also garbage collection routines, term-copying,
and database modification predicates) can be effectively implemented.
Rational trees have been explicitly used in
graphics~\cite{eggert83:lp_graphics}, parser generation and grammar
manipulation~\cite{Colmerauer82,giannesini84:parser_generation},
computing with finite-state automata~\cite{Colmerauer82}, and several
areas of natural language processing: rational trees are used in
implementations of the HPSG (Head-driven Phrase Structure Grammar)
formalism~\cite{pollard84:HPSG}, in the ALE (Attribute Logic Engine)
system~\cite{carpenter92:typed_feature_structures}, and in the ProFIT
(Prolog with Features, Inheritance and Templates)
system~\cite{erbach95:profit}.

In this paper we present yet another application of rational trees:
the implementation of interpreters for procedural languages featuring
explicit control instructions (such as \texttt{jump}s,
\texttt{if-then-else}s, \texttt{while}s and \texttt{repeat}s) which
can divert the execution flow.  Our aim is to implement an interpreter
which is simple and easy to verify, but without sacrificing
performance unnecessarily.  Also, with the aim of producing an
interpreter amenable to be analyzed and transformed, we want it to be
as clean and trick-free possible.  Some of the techniques we present
can be performed using attribute variables~\cite{holzbaur-plilp92} and
occurs-check unification, but the resulting program is less amenable
to automatic analysis and transformation. Describing it falls beyond
the scope of this paper.

In the rest of the paper we will first describe our initial problem
(Section~\ref{sec:target}) and then we will describe and evaluate two
solutions (Sections~\ref{sec:metainterpreter_free}
and~\ref{sec:enhanced}).  We include code for the interpreters
developed and for the benchmark programs.

\begin{table}[htbp]
  \centering
  \begin{tabular}{|l|p{0.7\textwidth}|}\hline
\texttt{label:instruction} & 
    \emph{\texttt{label} is the address of \texttt{instruction}} \\\hline
\texttt{load <number>} & 
    \emph{load \texttt{<number>} in the accumulator} \\\hline
\texttt{load <mem>} & 
    \emph{load the contents of the address \texttt{<mem>} in the  accumulator} \\\hline
\texttt{sto <mem>} & 
    \emph{store the contents of the accumulator in the address \texttt{<mem>}} \\\hline
\texttt{add <number>} & 
    \emph{add \texttt{<number>} to the accumulator and leave the
      result in the accumulator} \\\hline 
\texttt{add <mem>} & 
    \emph{add the contents of \texttt{<mem>} to the accumulator  and
      leave the result in the accumulator} \\\hline 
\texttt{sub <number>} & 
    \emph{subtract \texttt{<number>} from the accumulator and leave
      the result in the accumulator} \\\hline 
\texttt{sub <mem>} & 
    \emph{subtract the contents of the address \texttt{<mem>} from
      the accumulator and leave the result in the accumulator}
  \\\hline 
\texttt{jmp <label>} & 
    \emph{jump to the instruction at the address \texttt{<label>}} \\\hline 
\texttt{jez <label>} & 
    \emph{jump to the instruction at the address \texttt{<label>} if
      the accumulator is equal to zero} \\\hline 
\texttt{jnez <label>} & 
    \emph{jump to the instruction at the address \texttt{<label>} if
      the accumulator is not equal to zero} \\\hline 
\texttt{nop} & \emph{do nothing} \\\hline
  \end{tabular}
  \caption{Syntax and (short) semantics of a simple language}
  \label{tab:syntax-semantics}
\end{table}

\section{The Problem and the Target Language}
\label{sec:target}

We want to interpret a procedural language using a (high-level) logic
language (Prolog, in our case); there are several reasons to perform such
interpretation:

\begin{itemize}
\item We may want to execute programs in a language for which there is
  no compiler or interpreter available, and we do not want to invest
  time in writing a more sophisticated compiler.  A simple interpreter
  might be enough for our purposes.
  
\item We may want to experiment with alternative semantics for the
  same (syntactically speaking) language.  It is clearly much easier
  to do that if the interpreter is easy to understand and change.
  
\item We may want to apply analyzers written for the interpreter
  language to the interpreted language.  This is possible by first
  unfolding the interpreter with respect to the interpreted program,
  analyzing the resulting unfolded program, and then reinterpreting
  the analysis in the domain of the interpreted program.
\end{itemize}

We will first set down a toy target language which notwithstanding
features all the characteristics we need.  For simplicity we will
adopt an assembler-like language whose syntax and semantics are easy
to describe.  A summary of the (minimal) set of instructions we have
chosen is shown in Table~\ref{tab:syntax-semantics}.
Figures~\ref{fig:square},~\ref{fig:fibonacci}, and~\ref{fig:factorial}
show several programs in the target language (those we have used to
evaluate our implementations).

The target machine has a single register, the accumulator, which is
used in all but two instructions, and which acts as implicit source
and destination for data.  Explicit data sources are either numeric
constants (which appear directly in the code) or symbolic labels,
which represent locations where the operands are stored. Data labels
do not need to be declared, as the assembler / interpreter assigns
them a location.  Data is loaded from implicit operands or from
addresses into the accumulator; data in the accumulator can be stored
in memory addresses represented by labels.  Arithmetic operations
involve the accumulator both as source of values and as destination of
results.  The second operand for arithmetic instructions is explicit
in the program source.

Instructions can be labeled with a symbolic name which is used by
control instructions; an unbound number of such label names are
available.  These labels appear as operands of jumps; unconditional
jumps make the execution continue at the code associated to the label.
Conditional jumps test the accumulator value and decide whether to
continue executing on the next instruction of the program, or if
execution has to divert to the instruction whose label is the operand
of the \texttt{jump} instruction.  There is no state register (usual
in real microprocessors) which changes with the accumulator
operations.  There is also a no-op instruction.
Execution starts in the first instruction of the program, and finishes
when attempting to continue executing after the last instruction.
There is no \texttt{halt} instruction.\footnote{It may be replaced by
  making the last program instruction a \texttt{nop} and jumping to
  it.}

\begin{figure}[htbp]
  \centering
  \begin{minipage}{0.95\textwidth}
\small
\linespread{0.9}
\hrule
\medskip
\begin{verbatim}
program(fibo,
        [jnez(calculate), load(0), sto(curr), jmp(end), calculate:sto(ind),
         load(0), sto(prev), load(1), sto(curr), start_loop:load(ind), sub(1),
         sto(ind), jez(end), load(curr), sto(inter), add(prev), sto(curr),
         load(inter), sto(prev), jmp(start_loop), end:load(curr)]).
\end{verbatim}
\hrule
  \end{minipage}
  \caption{Prolog representation of program in Figure~\ref{fig:fibonacci}}
  \label{fig:prolog-fibonacci}
\end{figure}

We will assume that programs are already in the form of Prolog
terms.  Figure~\ref{fig:prolog-fibonacci} represents the Fibonacci
program in Figure~\ref{fig:fibonacci} as a fact, and every instruction
is assigned to a Prolog functor.
The interpreter and programs are designed to be used as
\begin{verbatim}
?- program(fibo, Program), interpret(Program, 400, A).
\end{verbatim}
where \texttt{400} is the initial value of the accumulator and
\texttt{A} is the contents of the accumulator at the end of the
execution.  By convention we take the contents of the accumulator at
the beginning of the execution as the input for the interpreted
program, and its value at the end as the output value of the program,
and we expect the programs to abide by this convention.  We also
assume that the programs are correct: for example, all labels to jump
to appear in the program, there are no repeated labels, etc.  The
interpreter will otherwise fail.

\begin{figure}[htbp]
\linespread{0.9}
\small
\centering
\begin{minipage}{0.9\textwidth}
\hrule
\medskip
\begin{verbatim}
interpret_nv(Program, AcIn, AcOut):-
        dic_empty(DicIn),
        interpret_naive(Program, Program, DicIn, _DicOut, AcIn, AcOut).

interpret_naive([], _Program, _, _, Acum, Acum).
interpret_naive([LI|Is], Program, DictIn, DictOut, AIn, AOut):-
        remove_label(LI, I),
        execute(I, Is, Program, NProgram, DictIn, DictMid, AIn, AMid),
        interpret_naive(NProgram, Program, DictMid, DictOut, AMid, AOut).

execute(load(NumOrLabel), Is, _Program, Is, D, D, _, Value):-
        eval(NumOrLabel, D, Value).
execute(add(NumOrLabel),  Is, _Program, Is, D, D, AcIn, AcOut):-
        eval(NumOrLabel, D, Value),
        AcOut is AcIn + Value.
execute(sub(NumOrLabel),  Is, _Program, Is, D, D, AcIn, AcOut):-
        eval(NumOrLabel, D, Value),
        AcOut is AcIn - Value.
execute(sto(Label),  Is, _Program, Is, DIn, DOut, Acum, Acum):-
        dic_replace(DIn, Label, Acum, DOut).
execute(jmp(Label), _, Program, Is, D, D, A, A):-
        Is = [Label:_|_], append(_, Is, Program).
execute(jez(Label), NIs, Program, Is, D, D, A, A):-
        (A = 0 -> Is = [Label:_|_], append(_, Is, Program) ; Is = NIs).
execute(jnez(Label), NIs, Program, Is, D, D, A, A):-
        (A \== 0 -> Is = [Label:_|_], append(_, Is, Program) ; Is = NIs).
execute(nop, Is, _Program, Is, D, D, A, A).

remove_label(_:I, I):- !.    eval(Number, _Dict, Number):-
remove_label(I, I).                  number(Number).
                             eval(Label, Dict, Number):-
                                     atom(Label),
                                     dic_get(Dict, Label, Number).
\end{verbatim}
\hrule
\end{minipage}
  \caption{An interpreter which does not need rational trees}
  \label{fig:simple-interpreter}
\end{figure}

\section{An Interpreter in the Herbrand Domain}
\label{sec:metainterpreter_free}

The two tasks the interpreter has to perform when executing the
instructions are maintaining the memory state (including the
accumulator) and deciding the next instruction to execute (and doing
so).  Conceptually, the memory is a mapping
\begin{displaymath}
\mathtt{label}  \mapsto \mathtt{value}
\end{displaymath}
which relates every memory address name to a single value.  This
mapping is updated with every program instruction which stores a value
into memory.  Its implementation can be left, as we will do, to a
table which takes the form of a Prolog dictionary, with the operations
in table~\ref{tab:table-operations}.

\begin{table}[htbp]
  \centering
  \begin{tabular}{|l|p{0.6\textwidth}|}\hline
    \texttt{dic\_empty(T)} & Unifies \texttt{T} with and empty table \\\hline
    \texttt{dic\_get(T,K,V)} & Unifies \texttt{V} with the value
    associated to the key \texttt{K} in table \texttt{T}; fails if
    \texttt{K} does not appear in \texttt{T}\\\hline
    \texttt{dic\_replace(TIn,K,V,TOut)} & Replaces the value associated to
    \texttt{K} in \texttt{TIn} for \texttt{V} and unify the new table
    with  \texttt{TOut}\\\hline
    \texttt{dic\_lookup(T,K,V)} & Unifies \texttt{V} with the value
    associated to the key \texttt{K} in table \texttt{T}, and inserts the
    association if it does not exist yet\\\hline
  \end{tabular}
  \caption{Table operations}
  \label{tab:table-operations}
\end{table}

There is a similar relationship between program labels and program
code: every program label maps into the sequence of instructions
starting at the labeled one.  With the current language, this mapping
does not change and is explicit in the program text, so it is possible
to search in the program text for the label to jump to when needed.
Figure~\ref{fig:simple-interpreter} portraits the complete code of an
interpreter for the target language, which assumes that programs are
already represented as in Figure~\ref{fig:prolog-fibonacci}.

\texttt{interpret\_nv/2} traverses the program executing its
instructions as they are found.  When an instruction which has to
store data in memory is found, the table which implements the memory
is updated.  The accumulator is explicitly passed around in a pair of
arguments, corresponding to its state before and after executing each
instruction.
It could have been assigned a (designated) location in the table which
represents the memory;
however, as the accumulator is used in almost all instructions, having
a direct access to it is reasonable and easy to code.
Jumping to a label is performed by searching the corresponding point
in the program.  This makes the jump operation to have complexity
$O(n)$, where $n$ is the size (in instructions) of the program.  This
overhead does not affect the run-time complexity of the interpreted
program, since the overhead is constant for every program, and does
not change with the input arguments.

We will not prove formally that this interpreter does not generate
rational trees, but we will give some intuition supporting this.  In
the \textbf{intended} use of \texttt{interpret\_nv/3}, the
\texttt{Program} input argument is ground and \texttt{AcIn} should
also be ground, and both should be Herbrand terms.  Therefore, in the
\texttt{interpret\_naive/6} predicate the first, second, and fourth
predicates should be ground.  All the operations take place in
\texttt{execute/8}, and in this predicate arguments 1 to 4 and 7 are
ground at call time.  We need of course the code of the dictionary
predicates to make sure that no cyclic terms are generated inside it,
but we can give this for granted with a suitable implementation.  The
dictionary is used only to store (ground) values associated to
(ground) keys and to retrieve them.  Thus, the only unifications which
take place in the interpreter have a ground term on one side of the
equation, and no cyclic data structures can be generated.

On the other hand it is straightforwar to make this program generate
rational trees with a \emph{non-intended} query: simply calling it as
\begin{verbatim}
?- interpret_nv([sto(where)], X, f(X)).
\end{verbatim}
will create a cyclic term.

\begin{figure}[htbp]
  \begin{minipage}{0.45\textwidth}
    \centering
    \includegraphics[width=\textwidth]{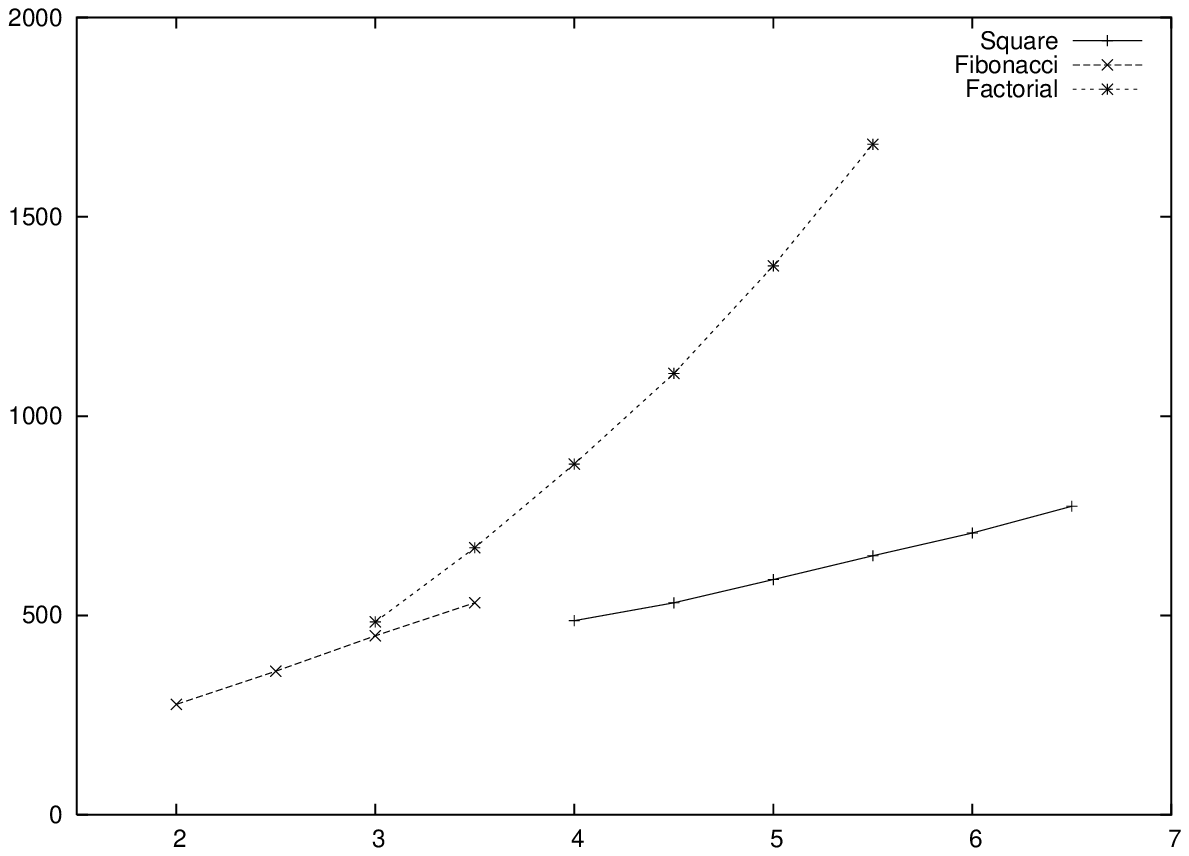}
    \caption{Time evolution for the interpreter free of occurs-check}
    \label{fig:occurs-check-free-graph}
  \end{minipage}
  \hfill
  \begin{minipage}{0.45\textwidth}
    \centering
    \includegraphics[width=\textwidth]{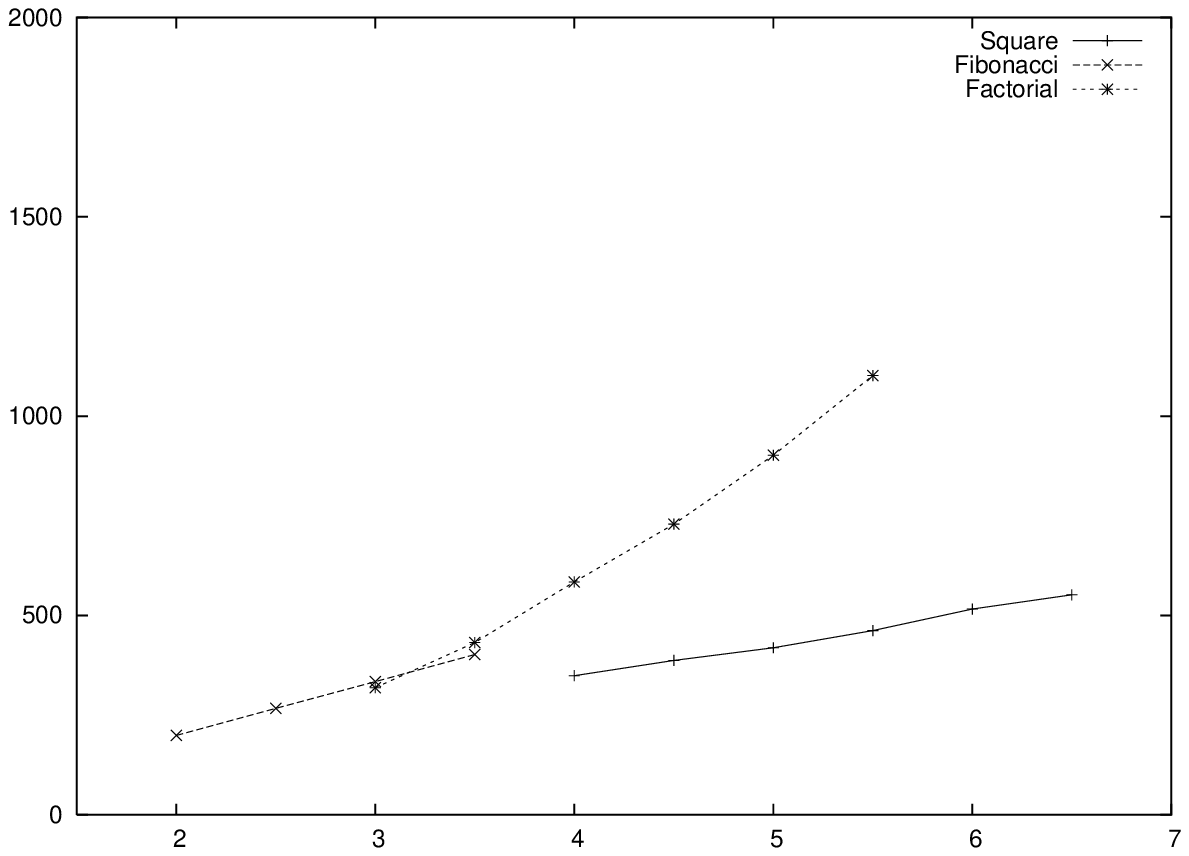}
    \caption{Time evolution for the rational tree interpreter}
    \label{fig:rational-graph}
  \end{minipage}
\end{figure}

Table~\ref{tab:exec-times-simple} shows the execution times for the
interpretation of the programs in figures~\ref{fig:square},
\ref{fig:fibonacci}, and \ref{fig:factorial}.  The same data is
depicted in Figure~\ref{fig:occurs-check-free-graph} (where the size
of the input data has been scaled to the range [1.5, 7.0] so that the
three graphs fit into the same figure).  Time is in milliseconds, and
was measured as the average of several runs.  That figure alone is not
really interesting, as it only shows the relative speed of different
programs which makes different calculations.

\begin{table}[htbp]
  \centering
  \begin{tabular}[t]{|c|r|} \hline
    \multicolumn{2}{|c|}{\textbf{\texttt{square}}} \\\hline
  \textbf{Input size} & \textbf{Time} \\\hline 
40000& 487 \\\hline
45000& 532 \\\hline
50000& 590\\\hline
55000& 650\\\hline
60000& 707\\\hline
65000& 774\\\hline
   \end{tabular}
  \begin{tabular}[t]{|c|r|} \hline
    \multicolumn{2}{|c|}{\textbf{\texttt{Fibonacci}}} \\\hline
  \textbf{Input size} & \textbf{Time} \\\hline 
20000& 277\\\hline
25000& 360\\\hline
30000& 449\\\hline
35000& 532\\\hline
   \end{tabular}
  \begin{tabular}[t]{|c|r|} \hline
    \multicolumn{2}{|c|}{\textbf{\texttt{factorial}}} \\\hline
 \textbf{Input size} & \textbf{Time} \\\hline %
 300&484 \\\hline
 350&670 \\\hline
 400& 880\\\hline
 450&1107\\\hline
 500&1377\\\hline
 550&1682\\\hline
   \end{tabular}
  \caption{Execution times, interpreter free of occurs-check}
  \label{tab:exec-times-simple}
\end{table}

\begin{figure}[htbp]
  \centering
\linespread{0.9}
\small
\begin{minipage}{0.85\textwidth}
\hrule
\medskip
\begin{verbatim}
interpret_thr(Program, AcIn, AcOut):-
        dic_empty(ProgramDicIn),
        thread_program(Program, ProgramDicIn, Threaded),
        dic_empty(MemDicIn),
        interpret_threaded(Threaded, MemDicIn, _MemDicOut, AcIn, AcOut).

thread_program([], _, end).
thread_program([Inst|Next], Dict, Threaded):-
        thread_instruction(Inst, Dict, Threaded, RestThreaded),
        thread_program(Next, Dict, RestThreaded).

thread_instruction(Label:Instruction, Dict, Threaded, RestThread):-
        dic_lookup(Dict, label:Label, Threaded),
        thread_instruction(Instruction, Dict, Threaded, RestThread).
thread_instruction(load(What), _,    load(What, Next), Next).
thread_instruction(sto(What),  _,    sto(What, Next), Next).
thread_instruction(add(What),  _,    add(What, Next), Next).
thread_instruction(sub(What),  _,    sub(What, Next), Next).
thread_instruction(nop,        _,    nop(Next),       Next).
thread_instruction(jmp(Label), Dict, jmp(Code),      _Next):-
        dic_lookup(Dict, label:Label, Code).
thread_instruction(jez(Label), Dict, jez(CodeYes, CodeNo), CodeNo):-
        dic_lookup(Dict, label:Label, CodeYes).
thread_instruction(jnez(Label), Dict,jnez(CodeYes, CodeNo), CodeNo):-
        dic_lookup(Dict, label:Label, CodeYes).
\end{verbatim}
\hrule
\end{minipage}
  \caption{Threading the code into a rational tree}
  \label{fig:threader}
\end{figure}

\begin{figure}[htbp]
  \centering
\linespread{0.9}
\small
\begin{minipage}{0.85\textwidth}
\hrule
\medskip
\begin{verbatim}
interpret_threaded(end, _, _, Acum, Acum).
interpret_threaded(Instr, DictIn, DictOut, AIn, AOut):-
        execute_threaded(Instr, Cont, DictIn, DictMid, AIn, AMid),
        interpret_threaded(Cont, DictMid, DictOut, AMid, AOut).

execute_threaded(load(NumOrLabel, Cont), Cont, D, D, _, Value):-
        eval(NumOrLabel, D, Value).
execute_threaded(add(NumOrLabel, Cont), Cont, D, D, AcIn, AcOut):-
        eval(NumOrLabel, D, Value), AcOut is AcIn + Value.
execute_threaded(sub(NumOrLabel, Cont), Cont, D, D, AcIn, AcOut):-
        eval(NumOrLabel, D, Value), AcOut is AcIn - Value.
execute_threaded(sto(Label, Cont),  Cont, DIn, DOut, Acum, Acum):-
        dic_replace(DIn, Label, Acum, DOut).
execute_threaded(jmp(Cont), Cont, D, D, A, A).
execute_threaded(jez(Yes, No), Cont, D, D, A, A):-
        (A = 0 -> Cont = Yes ; Cont = No).
execute_threaded(jnez(Yes, No), Cont, D, D, A, A):-
        (A \== 0 -> Cont = Yes ; Cont = No).
execute_threaded(nop(Cont), Cont, D, D, A, A).
\end{verbatim}
\hrule
\end{minipage}
  \caption{Interpreter for a rewritten program}
  \label{fig:enhanced-interpreter}
\end{figure}

\begin{table}[htbp]
  \centering
  \begin{tabular}[t]{|c|r|} \hline
    \multicolumn{2}{|c|}{\textbf{\texttt{square}}} \\\hline
  \textbf{Input size} & \textbf{Time} \\\hline 
40000& 349 \\\hline
45000& 387 \\\hline
50000& 419\\\hline
55000& 462\\\hline
60000& 516\\\hline
65000& 552\\\hline
   \end{tabular}
  \begin{tabular}[t]{|c|r|} \hline
    \multicolumn{2}{|c|}{\textbf{\texttt{Fibonacci}}} \\\hline
  \textbf{Input size} & \textbf{Time} \\\hline 
20000& 199\\\hline
25000& 267\\\hline
30000& 334\\\hline
35000& 402\\\hline
   \end{tabular}
  \begin{tabular}[t]{|c|r|} \hline
    \multicolumn{2}{|c|}{\textbf{\texttt{factorial}}} \\\hline
 \textbf{Input size} & \textbf{Time} \\\hline %
 300& 319 \\\hline
 350& 432 \\\hline
 400& 584\\\hline
 450&729\\\hline
 500&902\\\hline
 550&1102\\\hline
   \end{tabular}
  \caption{Execution times for the rational tree interpreter}
  \label{tab:exec-times-enhanced}
\end{table}

\begin{figure}[htbp]
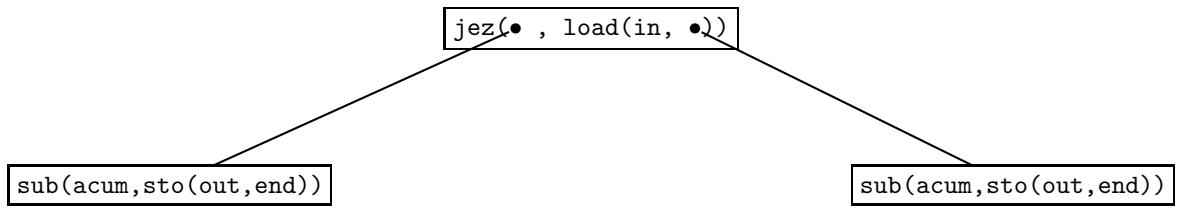

    \small
  \centering
\psmatrix
    & \fbox{\texttt{jez(\rnode{LNode}{$\bullet$} , load(in, \rnode{RNode}{$\bullet$}))}} & \\
\rnode{LLeaf}{\fbox{\texttt{sub(acum,sto(out,end))}}}
& & 
\rnode{RLeaf}{\fbox{\texttt{sub(acum,sto(out,end))}}}
\\
\endpsmatrix
\ncline{-}{LNode}{LLeaf}
\ncline{-}{RNode}{RLeaf}
  \caption{Tree structure for a threaded program with jumps forward}
  \label{fig:tree}
\end{figure}

\section{An Interpreter Which Uses Rational Trees}
\label{sec:enhanced}

The overhead brought about by the search of the code to jump to can be
diminished in several ways.  Performance can be improved by, for
example, storing program fragments in a table where labels are keys,
and  jumps would be preceded by a search for the correct fragment
of code:
\begin{verbatim}
execute(jmp(Label), _, Code, D, D, A, A):-
        dic_get(D, Label, Code).
\end{verbatim}
\texttt{Code} has had to be associated to \texttt{Label} in a previous
traversal of the program.  However, this code performs an unnecessary
search, since the representation of jump instructions (in fact, of
every instruction) can be augmented to contain the code to be executed
next, very similarly to a \emph{continuation passing} programming
style.  Conditional jump instructions need two addresses for the
\texttt{then} and the \texttt{else} branches, and the rest of the
instructions need just one.  
For example, the program fragment
\medskip

\centerline{\texttt{[sto(ind), load(0), sto(prev), load(1), sto(curr), load(ind)]}}
\medskip

\noindent
which does no have any jump instruction is translated by the predicate
\texttt{thread\_program/2} (Figure~\ref{fig:threader}) into \medskip

\centerline{\texttt{sto(ind,load(0,sto(prev,load(1,sto(curr,load(ind,end))))))}}
\medskip

\noindent
where the list structure is not needed any longer, since continuations
are explicit in the representation of the instructions, and the
pseudo-instruction \texttt{end} takes the place of the \texttt{[]}
ending the program.  Program fragments which have forward conditional
jumps generate trees: the code \medskip

\centerline{\texttt{[jez(is\_zero), load(in), is\_zero:sub(acum), sto(out)]}}
\medskip

\noindent
translates into the tree (Figure~\ref{fig:tree})
\medskip

\centerline{%
\texttt{jez(sub(acum,sto(out,end)),load(in,sub(acum,sto(out,end))))}}
\medskip

\noindent where the subtree \texttt{sub(acum,sto(out,end))} appears
twice and the labels are not needed any longer.\footnote{In any modern
  Prolog system these two subtrees will be internally represented
  physically as the same data structure, so there is no waste of
  memory.}

\begin{figure}[tbph]
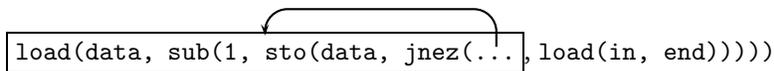

  \small
    \bigskip
  \centering
    \fbox{%
    \texttt{%
        load(data, sub(1, \pnode(0,0.15){OutBox}sto(data, jnez(\pnode(0.3,0){InBox}{...}%
            }%
    }, %
        \texttt{load(in, end)))))}
\ncangle[nodesep=5pt,angleA=90,angleB=90,linearc=0.3]{->}{InBox}{OutBox}
  \caption{Rational tree for a jump backwards}
  \label{fig:rational-tree}
\end{figure}

A code which performs a jump backward in the code, such as \medskip

\centerline{%
\texttt{[loop:load(data),sub(1),sto(data),jnez(loop),load(in)]}}
\medskip

\noindent
is translated into the rational tree (Figure~\ref{fig:rational-tree})
\medskip

\centerline{\texttt{load(data,sub(1,sto(data,jnez(load(data,sub(1,sto(...))),load(in,end)))))}}
\medskip

\noindent
where the dots abstract an infinite tree, but the whole term is a
rational tree.  A closer look at how threading is done reveals that
the unification which constructs the cyclic term is performed inside
the table implementation, in the first pass of the process which
builds an (intermediate) table associating labels with code.  A
trivial program such as \verb+loop:jmp(loop)+ generates a rational
tree due to the following table accesses and unifications (distilled
from the execution of the interpreter):
\begin{verbatim}
dic_lookup(Dict, loop, Threaded),
Threaded = jmp(Code),
dic_lookup(Dict, loop, Code)
\end{verbatim}
which actually executes \verb+T = jmp(C), T = C+ inside the code of
\verb+dic_lookup/3+.

In return for the first pass which threads the code constructing the
rational tree, the interpreter in
Figure~\ref{fig:enhanced-interpreter} is simpler than the one in
Figure~\ref{fig:simple-interpreter}, as no searches for the labels are
necessary: the code is directly accessible (in $O(1)$) in every
instruction.  This should bring performance advantages; this
assumption is supported by the numbers put forward in
Table~\ref{tab:exec-times-enhanced} and shown in
Figure~\ref{fig:enhanced-interpreter}, which are consistently better
than the ones for the previous interpreter.

\section{Conclusions}

We have shown two interpreters for an assembler language which differ
in the way jump instructions are handled.  The former performs linear
searches (which can however be improved), and the latter performs a
first pass which transforms the code into a rational tree which makes
it possible to execute jumps in $O(1)$.  We have evaluated the
performance of both interpreters, and the interpreter using rational
trees outperforms considerably the Herbrand-based one.

Therefore, using rational trees can bring benefits in terms of
performance, and supporting them correctly in Prolog systems should
make their use (and that of associated programming techniques) more
widespread.  However, giving full support is not straightforward, as
many builtins will have to be ready to accept and process them
correctly.  The programming style is also affected, because base cases
in recursion loops are not guaranteed to be reached, as they were with
(ground) Herbrand terms.  Additionally, there is no \emph{a priori}
way to detect, from the level of first order logic, that the predicate
is looping with current LP systems.

On the other hand, it does not seem to be a good idea to protect
indiscriminately library predicates against creation of cyclic terms,
for these predicates might be used to construct them on purpose, as in
our case.  It is true that hidden occurs checks can in some cases be
overcome if desired, by extracting culprit unifications, e.g., by
making
\begin{verbatim}
dic_lookup(Dict, loop, Threaded),
Threaded = jmp(Code),
dic_lookup(Dict, loop, ThreadedCode),
Code = ThreadedCode
\end{verbatim}
instead of the code shown in section~\ref{sec:enhanced}.  This kind of
transformations, which are correct if the called predicates are pure,
can however lead to termination (and correctness) problems in non-pure
predicates, and they have to be taken with a grain of salt.  We deem
much more useful to let the programmer to switch occurs-check on or
off if necessary, maybe aided or guided by the results of program
analysis.

\begin{figure}[htbp]
  \centering
  \linespread{0.9}
  \begin{minipage}[b]{0.45\textwidth}
\hrule
\texttt{
\begin{tabular}{rll}
& sto  & accum \\
& load & 1 \\
& sto & factor \\
& load & 0 \\
& sto & result \\
& load & accum \\
& jez & end \\
& sto & ind \\
loop:&load & result \\
& add & factor \\
& sto & result \\
& load & factor \\
& add & 2 \\
& sto & factor \\
& load & ind \\
& sub & 1 \\
& sto & ind \\
& jnez & loop \\
end: &load & result
\end{tabular}}
\hrule
  \caption{Squaring a number}
  \label{fig:square}
  \end{minipage}
    \hfill
  \begin{minipage}[b]{0.45\textwidth}
\hrule
\texttt{
  \begin{tabular}{rll}
&jnez&calculate\\
&load&0\\
&sto&curr\\
&jmp&end\\
calculate:&sto&ind\\
&load&0\\
&sto&prev\\
&load&1\\
&sto&curr\\
start\_loop:&load&ind\\
&sub&1\\
&sto&ind\\
&jez&end\\
&load&curr\\
&sto&inter\\
&add&prev\\
&sto&curr\\
&load&inter\\
&sto&prev\\
&jmp&start\_loop\\
end:&load&curr
  \end{tabular}
}
\hrule
  \caption{Code for the Fibonacci function}
  \label{fig:fibonacci}
  \end{minipage}
\end{figure}

\begin{figure}[htbp]
  \centering
\small
\linespread{0.9}
\begin{minipage}[t]{0.45\textwidth}
\hrule
\texttt{
  \begin{tabular}{rll}
&sto&accum\\
&load&1\\
&sto&res\\
&load&accum\\
&sub&1\\
&sto&n\\
&jez&exit\\
o\_loop:&load&res\\
&sto&add\\
&load&n\\
&sto&ind\\
&jez&dec\_idx\\
i\_loop:&load&res\\
&add&add\\
&sto&res\\
&load&ind\\
&sub&1\\
&sto&ind\\
&jnez&i\_loop\\
dec\_idx:&load&n\\
&sub&1\\
&sto&n\\
&jnez&o\_loop\\
exit:&load&res
\end{tabular}
}
\hrule
\end{minipage}
  \caption{Factorial program}
  \label{fig:factorial}
\end{figure}

\bibliographystyle{alpha}
\bibliography{/home/clip/bibtex/clip/clip,/home/clip/bibtex/clip/general,rational_trees}

\end{document}